\documentclass[prd,a4paper,twocolumn,amssymb,amsmath,nofootinbib,showpacs]{revtex4}
\newcommand{\dd}{\textrm{d\!I}}
\newcommand{\ww}{\wedge \kern-.75em \wedge}
\newcommand{\au}{
\begin{picture}(11,3)(-2,-2)
\put(2,5){$\scriptscriptstyle{1}$} \put(0,-3){A}
\end{picture}}

\newcommand{\ad}{
\begin{picture}(11,3)(-2,-2)
\put(2,5){$\scriptscriptstyle{2}$} \put(0,-3){A}
\end{picture}}

\begin{document}

\title{Diff-invariant Kinetic Terms in Arbitrary Dimensions}

\author{J. Fernando \surname{Barbero G.}}
%\surname{Barbero G.}
\email[]{jfbarbero@imaff.cfmac.csic.es}
\affiliation{Instituto de Matem\'aticas y F\'{\i}sica Fundamental,
C.S.I.C., C/ Serrano 113bis, 28006 Madrid, Spain}
\author{Eduardo J. \surname{S. Villase\~nor}}
%\surname{Villase\~nor}
\email[]{ejesus.sanchez@mat.ind.uem.es}
\affiliation{Escuela Superior de Ingenier\'{\i}a Industrial,
Universidad Europea, Urb. El Bosque, C/ Tajo s/n, Villaviciosa de
Od\'on, Madrid, 28670, Spain}

\date{March 25, 2002}

\begin{abstract}
We study the physical content of quadratic diff-invariant
Lagrangians in arbitrary dimensions by using covariant symplectic
techniques. This paper extends previous results in dimension four.
We discuss the difference between the even and odd dimensional
cases.
\end{abstract}


\maketitle

\section{\label{Intro}Introduction}
In recent years an interesting program devoted to the systematic
study of deformations of gauge theories has been developed
\cite{Barnich:1994pa}. The purpose of this line of research is to
study quantum field theories ``continuously connected" to other,
usually simpler, ones. In a typical situation one considers a free
theory, such as Maxwell electromagnetism, takes one or more copies
of the Lagrangian describing it, and studies the consistent
introduction of interaction terms \cite{Barnich:1994pa}. By doing
this it has been shown \cite{Henneaux:1997bm} that Yang-Mills
theories are the only consistent way to introduce interactions in
the Maxwell Lagrangian. However, this does not prove that these
are the only consistent interacting theories for 1-form fields
because for this to be true one must first know for sure that the
only possible starting point is the Maxwell action
\cite{Barbero:2000sb}.

This example suggests the problem of classifying all the free
theories with the scope of using them as a way to obtain
interesting interacting models. With this idea in mind, and with
the purpose of studying the possibility of finding suitable
diff-invariant kinetic terms that could be used to build
perturbative treatable gravitational actions we studied in
\cite{Barbero:1999ts} all the possible diff-invariant kinetic
terms in four space-time dimensions. We found out that these
theories do not describe any local degrees of freedom and, hence,
their consistent deformations cannot describe any gravitational
theory with local degrees of freedom. Anyway, having a full
classification of these kinetic terms one can consider to study
all their possible deformations as a means to classify the
topological theories in four dimensions of Schwarz type (see
\cite{Schwarz:2000cb} and references therein)\footnote{Topological
theories of the Witten type involve background structures, such as
metrics, and are outside the scope of this paper.} (or, at least,
a big subset of them). Here we extend the analysis of
\cite{Barbero:1999ts} to arbitrary dimensions. Even and odd
dimensional spacetimes show some diferences worth of study; in
particular in odd dimensions the kinetic term can have ``diagonal"
parts (as those appearing in the Chern-Simons Lagrangian) that are
not present in the even dimensional situation; we discuss this,
and related issues here. The paper is organized as follows. After
this introduction we study in section \ref{even} the physical
content of purely quadratic, diff-invariant Lagrangians in even
dimensions completing the results of \cite{Barbero:1999ts}. We
will do it by resorting to covariant symplectic methods as in the
quoted paper. Section \ref{odd} will discuss the odd dimensional
case with special emphasis on the differences with the even
dimensional one. We end the paper in section \ref{Concl} with our
conclusions and directions for future work.

\section{\label{even} Diff-invariant kinetic terms: The even
dimensional case}

We start by writing the most general diff-invariant local
quadratic action in an even-dimensional differentiable manifold.
Diff invariance demands the absence of background structures in
the Lagrangian, i.e. all the objects that appear must be taken as
dynamical. This precludes the appearance of background metrics or
connections (other than the ones defined directly by the
differential structure). The condition that the action be
quadratic also constraints its form in a twofold way because the
only derivative operator that we can use is the exterior
differential  and the only fields that can appear are differential
forms, acted upon by the exterior differential. If one considers
the inclussion of other types of tensor fields\footnote{One can
trivially add terms quadratic in tensor densities \emph{without
any derivatives} that do not describe any physical degrees of
freedom.} one would be forced to use covariant derivatives that
would give rise to non-quadratic terms in addition to the
quadratic ones (or include these terms as total divergencies that
would drop in the absence of boundaries).

Let us consider a $D+1$-dimensional differentiable manifold
$\mathcal{M}=\mathbb{R}\times\Sigma_D$ where $\Sigma_D$ is a
compact, orientable, $D$-dimensional manifold without boundary
($D=2N-1$, $N\in\mathbb{N}$). As we stated before we will work
with differential forms $\{A_n\}_{n=0}^{D+1}$,
$A_n\in\Omega_n^{\textrm{range}_n}(\mathcal{M})$. Here $A_n$
denotes a set of $n$-forms labelled by an internal index (taking
values from 1 to $\textrm{range}_n$) that we do not make explicit;
in practice we will take $A_n$ as a column vector with transpose
denoted as $A_n^{\rm{t}}$. The most general action under the
conditions expressed above is
\begin{widetext}
\begin{equation}
S_{2N}[A]=\int_{\mathcal{M}}\bigg\{\sum_{n=0}^{N-1}d\au_n^{\rm{t}}
\wedge\au_{2N-n-1}+\sum_{n=0}^{N-1}A_n^{\rm{t}}\wedge\Theta_nA_{2N-n}+
\frac{1}{2}A_N^{\rm{t}}\wedge\Theta_NA_N+
\theta^{\rm{t}}A_{2N}\bigg\}\,.\label{001}
\end{equation}
\end{widetext}
We discuss here the meaning and structure of the different terms
in (\ref{001}). The first term is the only one that involves
derivatives. In the following we will make a distinction between
two types of fields that we denote $\au$ and $\ad$; $\au$ are
those acted upon by derivatives (either directly or after
integration by parts) whereas no derivatives act on the $\ad$
fields. Notice that as we have the freedom to integrate by parts
we can extend the sum in the first term only to $N-1$. Another
interesting feature of this first term is that, as $n\neq 2N-n-1$
for all values of $n,\,N\in\mathbb{N}$, the two fields that appear
in it must be necessarily different so, as shown in appendix
\ref{redef}, it is actually possible to avoid the introduction of
a ``coupling matrix" and take $\au_n$ and $\au_{2N-n-1}$ with the
same number of internal components. This allows us to write it in
the following convenient form
$$
d\au_n^{\rm{t}}\wedge \au_{2N-n-1}= \left[
\begin{array}{ll}
d\au_n^{\rm{t}}& d\ad_n^{\rm{t}}
\end{array}
\right]\wedge \stackrel{1}{\Pi}\left[
\begin{array}{c}
\au_{2N-n-1}\\
\quad\\
\ad_{2N-n-1}
\end{array}
\right]
$$
with $\stackrel{1}{\Pi}:=\left[
\begin{array}{ll}
 \mathbb{I}_n& 0\\
0 & 0
\end{array}
\right]$ ($\mathbb{I}_n$ is the $n\times n$ identity matrix). The
second term involves both type 1 and type 2 fields and it does not
contain any derivatives. We have used the following compact
notation
$$
A_n^{\rm{t}}\wedge\Theta_nA_{2N-n} \hspace{-1mm}=
\hspace{-1mm}\left[
\begin{array}{ll}
\au_n^{\rm{t}}& \ad_n^{\rm{t}}
\end{array}
\right]\wedge \left[
\begin{array}{ll}
\Theta^{11}_n& \Theta^{12}_n\\
\quad & \quad \\
\Theta^{21}_n& \Theta^{22}_n
\end{array}
\right] \hspace{-2mm}\left[
\begin{array}{c}
\au_{2N-n}\\
\quad \\
\ad_{2N-n}
\end{array}
\right].
$$
Here we must allow the appearance of arbitrary, real, coupling
matrices $\Theta^{11}_n$, $\Theta^{12}_n$, $\Theta^{21}_n$, and
$\Theta^{22}_n$. These matrices are not square, in general,
because the dimensions of $A_n$ and $A_{2N-n}$ do not necessarily
match. We are also writing type-1 and type-2 fields in a single
object
$$
A_n=\left[
\begin{array}{c}
\au_{n}\\
\quad \\
\ad_{n}
\end{array}\right]\,.
$$

The third term in (\ref{001}) involves only the $A_N$ field so the
matrix $\Theta_N$ is symmetric or antisymmetric for even and odd
values of $N$ respectively; for convenience we introduce a 1/2
factor in front of it. Finally the last, linear term, involves a
2N-form $A_{2N}$ and a constant vector $\theta$.

The field equations obtained from (\ref{001}) are
\begin{subequations}
\label{002}
\begin{eqnarray}
& &\hspace{-1.3cm}\begin{array}{ll} \Theta_0^{\rm{t}}
A_0+\theta=0,
\end{array}\label{002a}\\
& &\hspace{-1.3cm}\begin{array}{ll}
\stackrel{1}{\Pi}dA_n+\Theta_{n+1}^{\rm{t}}A_{n+1}=0,
\hspace{1.7cm}0\leq n<N-1
\end{array}\label{002b}\\
& &\hspace{-1.3cm}\begin{array}{ll}
(-1)^N\stackrel{1}{\Pi}dA_{N-1}+ \Theta_{N}A_{N}=0, &
\end{array}\label{002c}\\
& &\hspace{-1.3cm}\begin{array}{ll} (-1)^n\stackrel{1}{\Pi}dA_n+
\Theta_{2N-n-1}A_{n+1}=0, \quad N\leq n< 2N.
\end{array}\label{002d}
\end{eqnarray}
\end{subequations}
Equation (\ref{002c}) will be referred to as the ``central
equation" because it plays a special role in the resolution of the
equations (\ref{002}).

The process of solving these equations is tedious but
straightforward. At every step, i.e. for every equation, we get a
certain consistency condition, a partial solution for one of the
fields, and a complete solution for another. The only obvious
exceptions are the first and last equations. For (\ref{002a}) we
do not get a complete solution for any of the $A_n$ whereas for
(\ref{002d}) the solution process terminates and we get no partial
solutions.

\subsection{\label{Eq1} Equation (\ref{002a})}

Let us take a set of linearly independent elements of the kernel
of $\Theta_0$, $\rho_{c_o}^{|0|}$ labelled by $c_0$
($\Theta_0\rho_{c_o}^{|0|}=0$). We immediately see that the
constant vector $\theta$ must satisfy the consistency condition
\begin{equation}
\rho_{c_o}^{|0|\rm{t}}\theta=0\label{cond01}.
\end{equation} When this
condition holds we can solve the linear set of equations
(\ref{002a}) to get
\begin{equation}
A_0(x)=-\Theta_0^{-1\rm{t}}\theta+
\lambda_{\alpha_0}^{|0|}A_0^{\alpha_0}(x)\label{cond02}
\end{equation}
where $A_0^{\alpha_0}(x)$ are \emph{arbitrary} 0-forms at this
stage, $\lambda_{\alpha_0}^{|0|}$, labelled by $\alpha_0$, belong
to the kernel of $\Theta_0^{\rm{t}}$
($\Theta_0^{\rm{t}}\lambda_{\alpha_0}^{|0|}=0$), and
$\Theta_0^{-1\rm{t}}\theta$ denotes a particular solution to the
inhomogeneous equations (\ref{002a})\footnote{Upper and lower
indices appearing in pairs are summed over. In order to avoid
confusion we use vertical bars in the labels of $\rho$ and
$\lambda$ that indicate to which of the $\Theta$ matrices they
refer to; this indices are \emph{never} summed over.
$\Theta_0^{-1}$ can be thought of as a pseudoinverse of
$\Theta_0$.}. At this point we have a consistency condition
(\ref{cond01}) and a partial solution for $A_0(x)$ that we use in
the next step.

\subsection{\label{Eq2} Equations (\ref{002b})}

If $N>1$ Eqs. (\ref{002b}) are a set of $N-1$ equations that are
all solved by following essentially the same procedure as before
though there are some minor differences depending on the type of
differential  forms involved. Let us consider first the equation
for $n=0$,
$$
\stackrel{1}{\Pi}dA_0+\Theta_{1}^{\rm{t}}A_{1}=0\,.
$$
Taking a set of linearly independent elements of the kernel of
$\Theta_1$, $\rho_{c_1}^{|1|}$ labelled by $c_1$
($\Theta_1\rho_{c_1}^{|1|}=0$) we obtain the consistency condition
\begin{equation}
\rho_{c_1}^{|1|\rm{t}}\stackrel{1}{\Pi}
\lambda_{\alpha_0}^{|0|}dA_0^{\alpha_0}(x):=
\mathcal{M}_{c_1\alpha_0}^{|0|}dA_0^{\alpha_0}(x)=0\label{cond11}
\end{equation}
where we have defined a new matrix
$\mathcal{M}_{c_1\alpha_0}^{|0|}$. Notice that the presence of
$\stackrel{1}{\Pi}$ in its definition means that (\ref{cond11}) is
a condition on type-1 fields only. This condition can be solved in
the following way. We expand
$$
A_0^{\alpha_0}(x)=[r^{|0|}_{p_0}]^{\alpha_0}A_0^{p_0}(x)+
[v^{|0|}_{q_0}]^{\alpha_0} A_0^{q_0}(x)
$$
where
$\mathcal{M}_{c_1\alpha_0}^{|0|}[r^{|0|}_{p_0}]^{\alpha_0}=0$ and
the $[v^{|0|}_{q_0}]^{\alpha_0}$ together with
$[r^{|0|}_{p_0}]^{\alpha_0}$ define a basis of the real vector
space on which $\mathcal{M}_{c_1\alpha_0}^{|0|}$ is defined. Here
$p_0$, $q_0$ are supposed to take different sets of values (so
that $A^{p_0}(x)$ and $A^{q_0}(x)$ are independent objects) and
label the vectors in the basis; $\alpha_0$ is an explicit vector
index. Plugging the previous decomposition in (\ref{cond11}) gives
\begin{equation}
\mathcal{M}_{c_1\alpha_0}^{|0|}[v^{|0|}_{q_0}]^{\alpha_0}
dA_0^{q_0}(x)=0\,,\label{ecu homo}
\end{equation}
and taking into account that
$\mathcal{M}_{c_1\alpha_0}^{|0|}[v^{|0|}_{q_0}]^{\alpha_0}$ is a
basis of the image of $\mathcal{M}_{c_1\alpha_0}^{|0|}$ we
conclude that $dA_0^{q_0}(x)=0$ and, hence
\begin{equation}
A_0^{q_0}(x)=\mathbf{a}_{|0|}^{q_0i_0}A_{i_0}^{|0|}(x)\,,\label{cond12}
\end{equation}
where $\{A_{i_0}^{|0|}(x)\}_{i_0=0}^{\dim H^0(\mathcal{M})}$ is a
basis of the $0^{\textrm{th}}$-de Rham cohomology
group\footnote{For a short review on the de-Rham cohomology groups
in this context we refer the reader to \cite{Barbero:1999ts}.} of
$\mathcal{M}$ and $\mathbf{a}_{|0|}^{q_0i_0}\in\mathbb{R}$. With
this information we can complete now the partial solution for
$A_0(x)$ that we obtained in the previous step
\begin{eqnarray}
&&\hspace{-.5cm}A_0(x)=\label{solA0}\\
&&-\Theta_0^{-1\rm{t}}\theta+\lambda_{\alpha_0}^{|0|}\big\{
[r^{|0|}_{p_0}]^{\alpha_0}A_0^{p_0}(x)+ [v^{|0|}_{q_0}]^{\alpha_0}
\mathbf{a}_{|0|}^{q_0i_0}A_{i_0}^{|0|}(x)\big\}\,.\nonumber
\end{eqnarray}
We can also get in this step, by using (\ref{solA0}), a partial
solution for $A_1$ in the form
\begin{equation}
A_1(x)=-\Theta_1^{-1\rm{t}}\stackrel{1}{\Pi}
\lambda_{\alpha_0}^{|0|}[r^{|0|}_{p_0}]^{\alpha_0}dA_0^{p_0}(x)+
\lambda_{\alpha_1}^{|1|}A_1^{\alpha_1}(x)\,.\label{solparA1}
\end{equation}
The remaining equations in (\ref{002b}) are solved in exactly the
same way. The only difference with the case just discussed appears
in the solutions of conditions analogous to (\ref{ecu homo}) for
$p$-forms with $p>0$. If we have, for example,
$$
\mathcal{M}_{c_2\alpha_1}^{|1|}[v^{|1|}_{q_1}]^{\alpha_1}
dA_1^{q_1}(x)=0
$$
with $\mathcal{M}_{c_2\alpha_1}^{|1|}:=
\rho_{c_2}^{|2|\rm{t}}\stackrel{1}{\Pi}\lambda_{\alpha_1}^{|1|}$
we expand
$$
A_1^{\alpha_1}(x)=[r^{|1|}_{p_1}]^{\alpha_1}A_1^{p_1}(x)+
[v^{|1|}_{q_1}]^{\alpha_1} A_1^{q_1}(x)
$$
where
$\mathcal{M}_{c_2\alpha_1}^{|1|}[r^{|1|}_{p_1}]^{\alpha_1}=0$ and
the $[v^{|1|}_{q_1}]^{\alpha_1}$ are introduced to complete a
basis of the real vector space on which
$\mathcal{M}_{c_2\alpha_1}^{|1|}$ is defined. By doing this we get
the equation $dA_1^{q_1}(x)=0$ with solutions given by
\begin{equation}
A_1^{q_1}(x)=\mathbf{a}_{|1|}^{q_1i_1}A_{i_1}^{|1|}(x)+d\varpi_0^{q_1}(x).
\label{solA1q}
\end{equation}
Here  $\{A_{i_1}^{|1|}(x)\}_{i_1=0}^{\dim H^1(\mathcal{M})}$ is a
basis of the $1^{\textrm{st}}$-de Rham cohomology group of
$\mathcal{M}$, $\mathbf{a}_{|1|}^{q_1i_1}\in\mathbb{R}$, and
$\varpi_0^{q_1}(x)$ are arbitrary 0-forms. The only difference
with the case discussed above is the appearance of
$\varpi_0^{q_1}(x)$. Similar objects appear for the remaining
equations. We end this subsection by giving the solutions obtained
from a generic equation (labelled by n) in this set: the complete
solution for $A_n$, $(0<n<N-1)$
\begin{eqnarray}
&&\hspace{-.5cm}A_n(x)=-\Theta_n^{-1\rm{t}}\stackrel{1}{\Pi}
\lambda_{\alpha_{n-1}}^{|n-1|}[r^{|n-1|}_{p_{n-1}}]^{\alpha_{n-1}}
dA_{n-1}^{p_{n-1}}(x)+\nonumber\\
&&\hspace{-.5cm}+\lambda_{\alpha_n}^{|n|}
\bigg\{[r^{|n|}_{p_n}]^{\alpha_n}A^{p_n}_n(x)+
\label{CompSolAn}\\
&&\hspace{2cm}+[v^{|n|}_{q_n}]^{\alpha_n}
\bigg[\mathbf{a}_{|n|}^{q_ni_n}A^{|n|}_{i_n}(x)+
d\varpi_{n-1}^{q_n}(x)\bigg]\bigg\}\nonumber
\end{eqnarray}
and a partial solution for $A_{n+1}$ $(0\leq n<N-1)$
\begin{eqnarray}
&&\hspace{-.8cm}A_{n+1}(x)=\label{solparAn+1}\\
&&\hspace{.4cm}-\Theta_{n+1}^{-1\rm{t}}\stackrel{1}{\Pi}
\lambda_{\alpha_n}^{|n|} [r^{|n|}_{p_n}]^{\alpha_n} dA_n^{p_n}(x)+
\lambda_{\alpha_{n+1}}^{|n+1|}A_{n+1}^{\alpha_{n+1}}(x)\,.\nonumber
\end{eqnarray}
\subsection{\label{Eq3} Central equation (\ref{002c})}
The matrix $\Theta_N$ that appears in this equation is either
symmetric or antisymmetric and hence the kernel of $\Theta_N$ and
$\Theta_N^{\rm{t}}$ coincide. Here we get the complete solution
for $A_{N-1}(x)$
\begin{eqnarray}
&&\hspace{-.5cm}A_{N-1}(x)=-\Theta_{N-1}^{-1\rm{t}}\stackrel{1}{\Pi}
\lambda_{\alpha_{N-2}}^{|N-2|}[r^{|N-2|}_{p_{N-2}}]^{\alpha_{N-2}}
dA_{N-2}^{p_{N-2}}(x)\nonumber\\
&&\hspace{-.5cm}+\lambda_{\alpha_{N-1}}^{|N-1|}
\bigg\{[r^{|N-1|}_{p_{N-1}}]^{\alpha_{N-1}}A^{p_{N-1}}_{N-1}(x)\label{solcompAN-1}\\
&&\hspace{.3cm}+[v^{|{N-1}|}_{q_{N-1}}]^{\alpha_{N-1}}
\big[\mathbf{a}_{|{N-1}|}^{q_{N-1}i_{N-1}}
A^{|{N-1}|}_{i_{N-1}}(x)+d\varpi_{N-2}^{q_{N-1}}(x)\big]\bigg\}
\nonumber
\end{eqnarray}
and a partial solution for $A_N(x)$
\begin{eqnarray}
&&\hspace{-.5cm}A_{N}(x)=(-1)^{N+1}\Theta_{N}^{-1}\!\!\stackrel{1}{\Pi}
\lambda_{\alpha_{N-1}}^{|N-1|}
[r^{|N-1|}_{p_{N-1}}]^{\alpha_{N-1}} dA_{N-1}^{p_{N-1}}(x)\nonumber\\
&& \hspace{.9cm}+\rho_{c_N}^{|N|}A_{N}^{c_{N}}(x).\label{solparAN}
\end{eqnarray}
\subsection{\label{Eq4} Equations (\ref{002d})}
At this point the procedure that we use to solve the equations
must be clear and, in fact, the rationale to give some details in
this and the following sections is that of introducing in a
systematic way the notation that we are using. Let us consider
first the equation ($n=N$)
$$
(-1)^N\stackrel{1}{\Pi}dA_N+ \Theta_{N-1}A_{N+1}=0.
$$
We have now the consistency condition given by
$$
\lambda_{\alpha_{N-1}}^{|N-1|\rm{t}}
\stackrel{1}{\Pi}\rho_{c_N}^{|N|}dA_N^{c_N}(x):=
\mathcal{N}_{\alpha_{N-1}c_N}^{|N|}dA_N^{c_N}(x)=0.
$$
where  ${\cal N}^{|k|}_{\alpha_{k-1}c_{k}}$ satisfies ${\cal
N}^{|k|}_{\alpha_{k-1}c_{k}}={\cal M}^{|k-1|}_{c_k\alpha_{k-1}}$
for every $k$. Expanding
$$
A_N^{c_N}(x)=[l_{p_N}^{|N|}]^{c_N}A_N^{p_N}(x)+
[w_{q_N}^{|N|}]^{c_N}A_N^{q_N}(x)
$$
with $\mathcal{N}_{\alpha_{N-1}c_N}^{|N|}[l_{p_N}^{|N|}]^{c_N}=0$
and $[w_{q_N}^{|N|}]^{\alpha_N}$ introduced to complete a basis of
the vector space where $\mathcal{N}_{\alpha_{N-1}c_N}^{|N|}$ is
defined. The complete solution for $A_N(x)$ is
\begin{eqnarray}
&&\hspace{-.5cm}A_N(x)=(-1)^{N+1}\Theta^{-1}_N
\stackrel{1}{\Pi}\lambda_{\alpha_{N-1}}^{|N-1|}
[r_{p_{N-1}}^{|N-1|}]^{\alpha_{N-1}}dA_{N-1}^{p_{N-1}}(x)\nonumber\\
&&\hspace{-.5cm}+\rho_{c_N}^{|N|}
\bigg\{[l_{p_N}^{|N|}]^{c_N}A_N^{p_N}(x)\label{SolCompAN}\\
&&\hspace{2cm}+[w_{q_N}^{|N|}]^{c_N}\bigg[
\mathbf{a}_{|N|}^{q_Ni_N}A_{i_N}^{|N|}(x)+
d\varpi_{N-1}^{q_N}(x)\bigg]\bigg\}\nonumber
\end{eqnarray}
and we get the folowing partial solution for $A_{N+1}(x)$
\begin{eqnarray}
&&\hspace{-.8cm}A_{N+1}(x)=(-1)^{N+1}\Theta_{N-1}^{-1}\stackrel{1}{\Pi}
\rho_{c_N}^{|N|} [l^{|N|}_{p_N}]^{c_N} dA_N^{p_N}(x)\nonumber\\
&&\hspace{1cm}+
\rho_{c_{N-1}}^{|N-1|}A_{N+1}^{c_{N-1}}(x).\label{solparAN+1}
\end{eqnarray}
From a generic equation in this set ($N<n<2N-1$) we get a complete
solution for $A_n(x)$
\begin{eqnarray}
&&\hspace{-.5cm}A_n(x)=\nonumber\\
&&\hspace{-.5cm}(-1)^n\Theta^{-1}_{2N-n}
\stackrel{1}{\Pi}\rho_{c_{2N-n+1}}^{|2N-n+1|}
[l_{p_{n-1}}^{|2N-n+1|}]^{c_{2N-n+1}}dA_{n-1}^{p_{n-1}}(x)\nonumber\\
&&\hspace{-.5cm}+\rho_{c_{2N-n}}^{|2N-n|}
\bigg\{[l_{p_n}^{|2N-n|}]^{c_{2N-n}}A_n^{p_n}(x)\label{SolCompAn}\\
&&\hspace{1cm}+[w_{q_n}^{|2N-n|}]^{c_{2N-n}}
\bigg[\mathbf{a}_{|n|}^{q_n i_n}A_{i_n}^{|n|}(x)+
d\varpi_{n-1}^{q_n}(x)\bigg]\bigg\}\nonumber
\end{eqnarray}
and a partial solution for $A_{n+1}(x)$
\begin{eqnarray}
&&\hspace{-.8cm}A_{n+1}(x)=\nonumber\\
&&\hspace{-.8cm}(-1)^{n+1}\Theta_{2N-n-1}^{-1}\stackrel{1}{\Pi}
\rho_{c_{2N-n}}^{|2N-n|} [l^{|2N-n|}_{p_n}]^{c_{2N-n}} dA_n^{p_n}(x)\nonumber\\
&&\hspace{-.8cm}+
\rho_{c_{2N-n-1}}^{|2N-n-1|}A_{n+1}^{c_{2N-n-1}}(x),\label{SolParAn+1}
\end{eqnarray}
where the objects appearing in these equations are defined in
analogy with the ones introduced at the beginning of this
subsection. The solutions for all the equations (\ref{002d}), are
given by (\ref{SolCompAn}) except the first one ($n=N$) which is
given by (\ref{SolCompAN}) and the last one that is given by

\begin{eqnarray}
&&\hspace{-.8cm}A_{2N}(x)=\label{solA2N}\\
&&\hspace{.5cm}\Theta_0^{-1}\stackrel{1}{\Pi}\rho_{c_1}^{|1|}
[l_{p_{2N-1}}^{|1|}]^{c_1}dA_{2N-1}^{p_{2N-1}}(x)+\rho_{c_0}^{|0|}A_{2N}^{c_0}(x).
\nonumber
\end{eqnarray}

\noindent As we see the solutions are parametrized by three
different types of objects: Arbitrary $n$-forms $A_n^{p_n}(x)$,
$(n=0,\ldots,2N-1)$, and $A_{2N}^{c_0}(x)$; arbitrary $n$-forms
$\varpi_n^{q_{n+1}}(x)$, $(n=0,\ldots,2N-2)$, and a set of real
numbers $\mathbf{a}_{|n|}^{q_ni_n}$ $(n=0,\ldots,2N-1)$.

If $N=1$ the solutions are given by (\ref{solA0}),
(\ref{SolCompAN}), and (\ref{solA2N}) with $N=1$ in these last two
equations.

\subsection{\label{symplectic} Symplectic structure}

Once we have obtained the solutions to the field equations we must
compute the symplectic structure on the solution space. This will
allow us to identify the physical degrees of freedom and the gauge
symmetries of the lagrangians introduced above. To this end we
must substitute the solutions to the field equations into the
symplectic structure obtained from the action
\begin{equation}
\Omega=\int_{\Sigma}\sum_{n=0}^{N-1}\dd\au_n^{\rm{t}}\ww\dd\au_{2N-n-1}
\label{omegaeven}
\end{equation} After a series of cancellations between the
different terms we find the following expression for the
symplectic structure in the space of fields
\begin{widetext}
\begin{equation}
\Omega=\sum_{n=0}^{N-1}\left[[v_{q_n}^{|n|}]^{\alpha_n}
\mathcal{N}_{\alpha_nc_{n+1}}^{|n+1|}[w_{q_{2N-n-1}}^{|n+1|}]^{c_{n+1}}
\int_{\Sigma} A_{i_n}^{|n|}\wedge
A_{i_{2N-n-1}}^{|2N-n-1|}\right]\dd \textbf{a}_{|n|}^{q_ni_n}\ww
\dd \textbf{a}_{|2N-n-1|}^{q_{2N-n-1}i_{2N-n-1}} \label{sympl}
\end{equation}
\end{widetext}
As we can see the only degrees of freedom are described by
$\textbf{a}_{|n|}^{q_ni_n}$; all the remaining objects in the
solutions obtained above represent gauge transformations. This
means that all these models (in even dimensional spacetimes) are
topological theories of the Schwarz type because they do not
depend on any metric but only on the topology of the manifold
$\mathcal{M}$ (or, equivalently $\Sigma$). Some comments on the
structure of (\ref{sympl}) are in order now. First it is
straightforward to prove that the factors in front of $\dd
\mathbf{a}_{|n|}^{q_ni_n}\ww \dd
\mathbf{a}_{|2N-n-1|}^{q_{2N-n-1}i_{2N-n-1}}$ are, in fact, non
singular. This is so because $
\mathcal{N}_{\alpha_nc_{n+1}}^{|n+1|}$ is non singular in the
vector spaces spanned by $[w_{q_{2N-n-1}}^{|n+1|}]^{c_{n+1}}$ and
$[v_{q_n}^{|n|}]^{\alpha_n}$ and the integral is non-singular as a
consequence of Poincar\'e duality. Second it is possible to prove,
by Poincar\'e's lemma that the de-Rham cohomology groups of
$\mathcal{M}$ and $\Sigma$ coincide
($H^k(\mathcal{M})=H^k(\Sigma)$). Finally we want to point out an
interesting feature of (\ref{sympl}) wich is the fact that every
term in the sum depends on two consecutive matrices $\Theta_n$ and
$\Theta_{n+1}$ and there are no ``diagonal terms" (involving
$\mathbf{a}_{|n|}^{q_ni_n}$'s with the same index $|n|$).

\section{\label{odd} Diff-invariant kinetic terms: The odd
dimensional case.}

Let us consider now the following action in $2N+1$ dimensions
\begin{widetext}
\begin{equation}
S_{2N+1}[A]=\int_{\mathcal{M}}\bigg\{\sum_{n=0}^{N-1}d\au_n^{\rm{t}}
\wedge\au_{2N-n}+\frac{1}{2}d\au_N\wedge\Xi_N\au_N+
\sum_{n=0}^{N}A_n^{\rm{t}}\wedge\Theta_nA_{2N-n+1}+
\theta^{\rm{t}}A_{2N+1}\bigg\}\label{actodd}
\end{equation}
\end{widetext}
where the notation that we use is analogous to the one introduced
in the previous section; in particular we work now with sets of
differential forms $A_n$ (with an internal index that we do not
make explicit) of order $n=0,\ldots,2N+1$. We see that we have now
the possibility of having a diagonal derivative term (that we will
refer to as the $\Xi$-term) $\frac{1}{2}d\au_N\wedge\Xi_N\au_N$.
At variance with the previous case we cannot eliminate the
coupling matrix $\Xi_N$ by non-singular field redefinitions. With
the conventions that we are using we can take $\Xi_N$ as a
non-singular matrix; if $N$ is odd we can choose it to be
symmetric whereas for even $N$ it is antisymmetric. If $\Xi_N$ is
symmetric and non-singular we can diagonalize $\Xi_N$  and write
it as a diagonal matrix of 1 and $-$1 entries. If it is
antisymmetric it is always possible to write it as the matrix
$$
\left[
\begin{array}{rr}
0 & \mathbb{I}\\
-\mathbb{I} & 0
\end{array}
\right]\,.
$$
This structure will determine, in part, the types of internal
symmetries of the actions derived from (\ref{actodd}) by means of
consistent deformations. The structure of the remaining terms of
this action is analogous to the ones considered in the previous
section. The field equations derived from (\ref{actodd}) are
\begin{subequations}
\label{oddeqs}
\begin{eqnarray}
& &\hspace{-1.3cm}\begin{array}{ll} \Theta_0^{\rm{t}}
A_0+\theta=0,
\end{array}\label{oddeqsa}\\
& &\hspace{-1.3cm}\begin{array}{ll}
\stackrel{1}{\Pi}dA_n+\Theta_{n+1}^{\rm{t}}A_{n+1}=0,
\hspace{1.7cm}0\leq n\leq N-1
\end{array}\label{oddeqsb}\\
& &\hspace{-1.3cm}\begin{array}{ll}
(-1)^{N+1}\Xi_N\stackrel{1}{\Pi}dA_{N}+\Theta_{N}A_{N+1}=0, &
\end{array}\label{oddeqsc}\\
& &\hspace{-1.3cm}\begin{array}{ll}
(-1)^{n+1}\!\!\stackrel{1}{\Pi}\!dA_n\!+\Theta_{2N-n}A_{n+1}\!=\!0,
\,\,\,N+1\!\leq n\leq \!2N.
\end{array}\label{oddeqsd}
\end{eqnarray}
\end{subequations}
The structure of equations (\ref{oddeqsa}-\ref{oddeqsb}) is
exactly the same as in the previous case; we write their solutions
here for completeness. From (\ref{oddeqsa}) we get the partial
solution

\begin{equation}
A_0(x)=-\Theta_0^{-1\rm{t}}\theta+
\lambda_{\alpha_0}^{|0|}A_0^{\alpha_0}(x)\label{condodd01}
\end{equation}
with $ \rho_{c_o}^{|0|\rm{t}}\theta=0$; and proceeding as before
we finally get
\begin{eqnarray*}
&&\hspace{-.5cm}A_0(x)=\nonumber\\
&&-\Theta_0^{-1\rm{t}}\theta+\lambda_{\alpha_0}^{|0|}\big\{
[r^{|0|}_{p_0}]^{\alpha_0}A_0^{p_0}(x)+ [v^{|0|}_{q_0}]^{\alpha_0}
\mathbf{a}_{|0|}^{q_0i_0}A_{i_0}^{|0|}(x)\big\}\,.\nonumber
\end{eqnarray*}
Equations (\ref{oddeqsb}) give the complete solution for $A_n$
($n=1,\ldots,N-1$)

\begin{eqnarray}
&&\hspace{-.5cm}A_n(x)=-\Theta_n^{-1\rm{t}}\stackrel{1}{\Pi}
\lambda_{\alpha_{n-1}}^{|n-1|}[r^{|n-1|}_{p_{n-1}}]^{\alpha_{n-1}}
dA_{n-1}^{p_{n-1}}(x)+\nonumber\\
&&\hspace{-.5cm}+\lambda_{\alpha_n}^{|n|}
\bigg\{[r^{|n|}_{p_n}]^{\alpha_n}A^{p_n}_n(x)+
\label{CompSoloddAn}\\
&&\hspace{2cm}+[v^{|n|}_{q_n}]^{\alpha_n}
\bigg[\mathbf{a}_{|n|}^{q_ni_n}A^{|n|}_{i_n}(x)+
d\varpi_{n-1}^{q_n}(x)\bigg]\bigg\}\nonumber
\end{eqnarray}

\noindent and a partial solution for $A_{n+1}$ ($n=0,\ldots,N-1$)

\begin{eqnarray}
&&\hspace{-.8cm}A_{n+1}(x)=\label{solparoddAn+1}\\
&&\hspace{.4cm}-\Theta_{n+1}^{-1\rm{t}}\stackrel{1}{\Pi}
\lambda_{\alpha_n}^{|n|} [r^{|n|}_{p_n}]^{\alpha_n} dA_n^{p_n}(x)+
\lambda_{\alpha_{n+1}}^{|n+1|}A_{n+1}^{\alpha_{n+1}}(x)\nonumber
\end{eqnarray}
with all the algebraic objects appearing in these expressions
defined as in the even case.

\subsection{\label{Eqoddc} Central equation (\ref{oddeqsc})}

The first consistency condition that we get from the central
equation is
\begin{equation}
\lambda^{|N|{\rm{t}}}_{\alpha_N}\Xi_N\stackrel{1}{\Pi}\lambda^{|N|}_{\alpha_N^{\prime}}
dA_N^{\alpha_N^{\prime}}(x):=\mathbb{M}_{\alpha_N\alpha_N^{\prime}}
dA_N^{\alpha_N^{\prime}}(x)=0\,.\label{matrixmm}
\end{equation}
Expanding
$$
A_n^{\alpha_N}(x)=[r^{|N|}_{p_N}]^{\alpha_N}A_N^{p_N}(x)+
[v^{|N|}_{q_N}]^{\alpha_N}A_N^{q_N}(x)
$$
where
$\mathbb{M}_{\alpha_N\alpha_N^{\prime}}[r^{|N|}_{p_N}]^{\alpha_N^{\prime}}=0$
and $[v^{|N|}_{q_N}]^{\alpha_N}$ are introduced, as in previous
instances, to complete a basis of the relevant vector space. The
complete solution for $A_N(x)$ is
\begin{eqnarray}
&&\hspace{-.5cm}A_N(x)=-\Theta_n^{-1\rm{t}}\stackrel{1}{\Pi}
\lambda_{\alpha_{N-1}}^{|N-1|}[r^{|N-1|}_{p_{N-1}}]^{\alpha_{N-1}}
dA_{N-1}^{p_{N-1}}(x)+\nonumber\\
&&\hspace{-.5cm}+\lambda_{\alpha_N}^{|N|}
\bigg\{[r^{|N|}_{p_N}]^{\alpha_N}A^{p_N}_N(x)+
\label{CompSoloddAN}\\
&&\hspace{2cm}+[v^{|N|}_{q_N}]^{\alpha_N}
\bigg[\mathbf{a}_{|N|}^{q_Ni_N}A^{|N|}_{i_N}(x)+
d\varpi_{N-1}^{q_N}(x)\bigg]\bigg\}\nonumber
\end{eqnarray}
and we get the following partial solution for $A_{N+1}(x)$
\begin{eqnarray}
&&\hspace{-.5cm}A_{N+1}(x)=\label{solparoddAN+1}\\
&&\hspace{-.2cm}(-1)^N\Theta_{N}^{-1}\Xi_N\stackrel{1}{\Pi}
\lambda_{\alpha_N}^{|N|} [r^{|N|}_{p_N}]^{\alpha_N} dA_N^{p_N}(x)+
\rho_{c_N}^{|N|}A_{N+1}^{c_N}(x)\,.\nonumber
\end{eqnarray}

\subsection{\label{Eqoddd} Equations (\ref{oddeqsd})}

\noindent Let us consider the equation corresponding to $n=N+1$
$$
(-1)^N\stackrel{1}{\Pi}dA_{N+1}+\Theta_{N-1}A_{N+2}=0\,.
$$
As before we get the consistency condition
$$
\lambda_{\alpha_{N-1}}^{|N-1|}\stackrel{1}{\Pi}
\rho_{c_N}^{|N|}dA_{N+1}^{c_N}(x):=
\mathcal{N}^{|N|}_{\alpha_{N-1}c_N}dA_{N+1}^{c_N}(x)=0\,.
$$
Expanding
$$
A_{N+1}^{c_N}(x)=[l^{|N|}_{p_{N+1}}]^{c_N}A^{p_{N+1}}_{N+1}(x)+
[w^{|N|}_{q_{N+1}}]^{c_N}A_{N+1}^{q_{N+1}}(x)
$$
with
$\mathcal{N}^{|N|}_{\alpha_{N-1}c_N}[l^{|N|}_{p_{N+1}}]^{c_N}=0$
and $[w^{|N|}_{q_{N+1}}]^{c_N}$ introduced to complete a basis of
the relevant vector space. The complete solution for $A_{N+1}$ is
\begin{eqnarray}
&&\hspace{-.5cm}A_{N+1}(x)=(-1)^N\Theta^{-1}_N\Xi_N
\stackrel{1}{\Pi}\lambda_{\alpha_N}^{|N|}
[r_{p_N}^{|N|}]^{\alpha_N}dA_N^{p_N}(x)\nonumber\\
&&\hspace{-.5cm}+\rho_{c_N}^{|N|}
\bigg\{[l_{p_{N+1}}^{|N|}]^{c_N}A_{N+1}^{p_{N+1}}(x)\label{SolCompoddAN+1}\\
&&\hspace{.5cm}+[w_{q_{N+1}}^{|N|}]^{c_N}\bigg[
\mathbf{a}_{|N+1|}^{q_{N+1}i_{N+1}}A_{i_{N+1}}^{|N+1|}(x)+
d\varpi_N^{q_{N+1}}(x)\bigg]\bigg\}\nonumber
\end{eqnarray}
and we get the following partial solution for $A_{N+2}$
\begin{eqnarray}
&&\hspace{-.8cm}A_{N+2}(x)=(-1)^{N+1}\Theta_{N-1}^{-1}\stackrel{1}{\Pi}
\rho_{c_N}^{|N|} [l^{|N|}_{p_{N+1}}]^{c_N} dA_{N+1}^{p_{N+1}}(x)\nonumber\\
&&\hspace{1cm}+
\rho_{c_{N-1}}^{|N-1|}A_{N+2}^{c_{N-1}}(x).\label{solparoddAN+2}
\end{eqnarray}
For a generic equation in this set ($N+1<n<2N$) we get a complete
solution for $A_n(x)$
\begin{eqnarray}
&&\hspace{-.5cm}A_n(x)=\nonumber\\
&&\hspace{-.5cm}(\!-1)^{n+1}\Theta^{-1}_{2N-n+1}
\!\!\stackrel{1}{\Pi}\!\rho_{c_{2N-n+2}}^{|2N-n+2|}
[l_{p_{n-1}}^{|2N-n+2|}]^{c_{2N-n+1}}dA_{n-1}^{p_{n-1}}(x)\nonumber\\
&&\hspace{-.5cm}+\rho_{c_{2N-n+1}}^{|2N-n+1|}
\bigg\{[l_{p_n}^{|2N-n+1|}]^{c_{2N-n+1}}A_n^{p_n}(x)\label{SolCompoddAn}\\
&&\hspace{.6cm}+[w_{q_n}^{|2N-n+1|}]^{c_{2N-n+1}}
\bigg[\mathbf{a}_{|n|}^{q_n i_n}A_{i_n}^{|n|}(x)+
d\varpi_{n-1}^{q_n}(x)\bigg]\bigg\}\nonumber
\end{eqnarray}
and a partial solution for $A_{n+1}(x)$
\begin{eqnarray}
&&\hspace{-.8cm}A_{n+1}(x)=\nonumber\\
&&\hspace{-.8cm}(-1)^n\Theta_{2N-n}^{-1}\stackrel{1}{\Pi}
\rho_{c_{2N-n+1}}^{|2N-n+1|} [l^{|2N-n+1|}_{p_n}]^{c_{2N-n+1}} dA_n^{p_n}(x)
\nonumber\\
&&\hspace{-.8cm}+
\rho_{c_{2N-n}}^{|2N-n|}A_{n+1}^{c_{2N-n}}(x),\label{SolParoddAn+1}
\end{eqnarray}
where the objects appearing in these equations are defined in
analogy with the ones previously introduced in the paper. The
solutions for all the equations (\ref{oddeqs}), are given by
(\ref{SolCompoddAn}) except the first one ($n=N+1$) which is given
by (\ref{SolCompoddAN+1}) and the last one that is given by
\begin{eqnarray}
&&\hspace{-.8cm}A_{2N+1}(x)=\label{soloddA2N+1}\\
&&\hspace{.5cm}\Theta_0^{-1}\stackrel{1}{\Pi}\rho_{c_1}^{|1|}
[l_{p_{2N}}^{|1|}]^{c_1}dA_{2N}^{p_{2N}}(x)+\rho_{c_0}^{|0|}A_{2N+1}^{c_0}(x).
\nonumber
\end{eqnarray}

\noindent The solutions are parametrized by three different types
of objects: Arbitrary $n$-forms $A_n^{p_n}(x)$, $(n=0,\ldots,2N)$,
and $A_{2N+1}^{c_0}(x)$; arbitrary $n$-forms
$\varpi_n^{q_{n+1}}(x)$, $(n=0,\ldots,2N-1)$, and a set of real
numbers $\mathbf{a}_{|n|}^{q_ni_n}$ $(n=0,\ldots,2N)$. If $N=1$
the solutions are
\begin{eqnarray*}
&&\hspace{-.3cm}A_0(x)=-\Theta_0^{-1\rm{t}}\theta\nonumber\\
&&\hspace{1.8cm}+\lambda^{|0|}_{\alpha_0}\bigg\{
[r_{p_0}^{|0|}]^{\alpha_0}A_0^{p_0}(x)+[v_{q_0}^{|0|}]^{\alpha_0}
\mathbf{a}_{|0|}^{q_0i_0}A_{i_0}^{|0|}(x)\bigg\}\nonumber\\
&&\hspace{-.3cm}A_1(x)=-\Theta_1^{-1\rm{t}}\stackrel{1}{\Pi}
\lambda^{|0|}_{\alpha_0}
[r_{p_0}^{|0|}]^{\alpha_0}dA_0^{p_0}(x)\nonumber\\
&&\hspace{-.3cm}+\lambda^{|1|}_{\alpha_1}\bigg\{
[r_{p_1}^{|1|}]^{\alpha_1}A_1^{p_1}(x)+[v_{q_1}^{|1|}]^{\alpha_1}\bigg[
\mathbf{a}_{|1|}^{q_1i_1}A_{i_1}^{|1|}(x)+d\varpi_0^{q_1}(x)\bigg]\bigg\}
\nonumber\\
&&\hspace{-.3cm}A_2(x)=-\Theta_1^{-1\rm{t}}\Xi_1\stackrel{1}{\Pi}
\lambda^{|1|}_{\alpha_1}
[r_{p_1}^{|1|}]^{\alpha_1}dA_1^{p_1}(x)\nonumber\\
&&\hspace{-.2cm}+\rho^{|1|}_{c_1}\bigg\{
[l_{p_2}^{|1|}]^{c_1}A_2^{p_2}(x)+[w_{q_2}^{|1|}]^{c_1}\bigg[
\mathbf{a}_{|2|}^{q_2i_2}A_{i_2}^{|2|}(x)+d\varpi_1^{q_2}(x)\bigg]\bigg\}
\nonumber\\
&&\hspace{-.3cm}A_3(x)=\Theta_0^{-1}\stackrel{1}{\Pi}
\rho_{c_1}^{|1|}[l^{|1|}_{p_2}]^{c_1}dA_2^{p_2}(x)+\rho_{c_0}^{|0|}A_3^{c_0}(x).
\noindent
\end{eqnarray*}

\subsection{\label{symplecticodd} Symplectic structure}

The symplectic structure in this case is obtained by substituting
the solutions to the field equations in
\begin{equation}
\Omega=\int_{\Sigma}\bigg[\sum_{n=0}^{N-1}\dd\au_n^{\rm{t}}\ww\dd\au_{2N-n}+
\dd\au_N^{\rm{t}}\ww\Xi_N\dd\au_{N} \bigg] \label{omegaodd}
\end{equation}
to get
\begin{widetext}
\begin{eqnarray}
&&\Omega=\sum_{n=0}^{N-1}\left[[v_{q_n}^{|n|}]^{\alpha_n}
\mathcal{N}_{\alpha_nc_{n+1}}^{|n+1|}[w_{q_{2N-n}}^{|n+1|}]^{c_{n+1}}
\int_{\Sigma} A_{i_n}^{|n|}\wedge A_{i_{2N-n}}^{|2N-n|}\right]\dd
\mathbf{a}_{|n|}^{q_ni_n}\ww \dd
\mathbf{a}_{|2N-n|}^{q_{2N-n}i_{2N-n}} \label{symplodd}\\
&&\quad\quad\quad+\left[[v_{q_N}^{|N|}]^{\alpha_N}
\mathbb{M}_{\alpha_N\alpha^{\prime}_N}
[v_{q^{\prime}_{N}}^{|N|}]^{\alpha^{\prime}_N}
\int_{\Sigma}A_{i_N}^{|N|}\wedge
A_{i_{N}^{\prime}}^{|N|}\right]\dd \mathbf{a}_{|N|}^{q_Ni_N}\ww
\dd \mathbf{a}_{|N|}^{q_{N}^{\prime}i_{N}^{\prime}}\,.\nonumber
\end{eqnarray}
\end{widetext}
The most interesting feature of $\Omega$ is the appearence of
diagonal $\dd\mathbf{a}{\wedge\kern-.55em\wedge}\dd\mathbf{a}$
terms that depend on both $\Xi_N$ and $\Theta_N$ (through
$\lambda_N$).

\section{\label{Concl}Concluding remarks}

We have studied in this paper the most general quadratic
diff-invariant theories in arbitrary dimensions; in particular we
have found that they always describe topological (non-local)
degrees of freedom. This result is easy to understand in the
covariant symplectic framework that we are using here to study the
dynamical content and the symmetries of the actions (\ref{001})
and (\ref{actodd}) because the symplectic structure $\Omega$ on
the solution space given by eqs. (\ref{sympl}) and
(\ref{symplodd}) is \emph{independent} of the choice of the
hypersurface $\Sigma$. As the functions that appear in the
solutions to the field equations are completely arbitrary the only
way to avoid a hypersurface dependence of $\Omega$ is if they do
not appear in the symplectic structure. This can be checked by the
explicit computation of the restriction of (\ref{omegaeven}) and
(\ref{omegaodd}) to the corresponding solution spaces.

In the actions that we have considered the fields that appear (or
rather, some combinations of them determined by the coupling
matrices that we introduce) can be thought of as Lagrange
multipliers imposing constraints on the exterior differentials of
other fields. The fact that the coupling matrices that we
introduce are completely general makes it difficult to disentangle
this complicated set of constraints. In fact, if instead of the
covariant symplectic techniques that we use here one resorts to
the more familiar Dirac constraint analysis, the problem becomes
very hard to solve. By finding out the most general solutions of
the field equations and the symplectic structure we can write down
the gauge symmetries of all these actions and identify their
degrees of freedom in a straightforward way.

As we can see by looking at the symplectic 2-forms that we find in
the paper the matrices that multiply the $\dd \mathbf{a} {\wedge
\kern-.55em \wedge}\dd \mathbf{a}$ terms depend on pairs of
consecutive $\Theta$ matrices and for odd dimensions also on $\Xi$
and $\Theta_N$.

An interesting and open problem is to study the consistent
deformations of the actions given in this paper (some work in this
direction has been carried out for the abelian Chern-Simons
Lagrangian in \cite{Barnich:1993vg}). They would provide coupled
topological theories of the BF type in even dimensions or coupled
Chern-Simons and BF theories in odd dimensions. One would expect
that some modes that are decoupled in the free models that we
consider here are actually coupled in their deformations as it
happens, in a slightly different context, in the BFYM Lagrangians
discussed in \cite{Cattaneo:1997fg}. For every Lagrangian of the
type that we are considering in this paper it is easy to find
another one involving only derivative terms with the same
dynamical content (changing, if necessary the internal dimensions
of some of the fields involved); however, we do not know if the
consistent deformations of Lagrangians equivalent in this sense
will be equivalent.

\appendix

\section{\label{redef}Structure of the derivative term in even dimensions}

We show here that, with the exception of the Chern-Simons-like
terms that appear in the odd-dimensional case (coupling $A_N$ with
$dA_N$) it is always possible to avoid introducing coupling
matrices in the derivative terms by using linear field
redefinitions.

If $\Delta\in{\cal M}_{_{M\times N}}(\mathbb{R})$ and we have
$dA^{\rm{t}}_m\wedge\Delta A_n$, we can introduce bases for
$\mathbb{R}^{N}$ and $\mathbb{R}^{M}$ as ${\cal
B}_n=\{v_{1},\dots,v_{r},\rho_{1},\dots,\rho_{N-r}\}\,$, ${\cal
B}_m=\{w_{1},\dots,w_{r},\lambda_{1},\dots,\lambda_{M-r}\}\,$
where $r={\rm rank}(\Delta)$, $\Delta \rho_{k}=0$ for
$k=1,\dots,N-r$, $\lambda_{j}^{\dagger}\Delta=0$ for
$j=1,\dots,M-r$. We have now
\begin{eqnarray}
dA^{\rm{t}}_m\wedge \Delta A_n= dA^{\rm{t}}_m({\cal
B}_m^{-1})^{\rm{t}} {\cal B}_m^{\rm{t}}\Delta{\cal B}_n {\cal
B}_n^{-1}A_n\nonumber\quad,
\end{eqnarray}
where ${\cal B}_m^{\rm{t}}\Delta{\cal B}_n $ has the following
block form
\begin{eqnarray}
\left[   \begin{array}{cc} w_{a}^{\rm{t}}\Delta v_{b}
&0\\0&0\end{array}\right] \nonumber\quad,
\end{eqnarray}
$w_{a}^{\rm{t}}\Delta v_{b} \in{\cal M}_{r\times r}(\mathbb{R})$
and is regular so that by independent linear redefinitions in
$A_m$ and $A_n$ it can be taken to be the identity. By using the
convention that fields that do no couple to derivatives are ``type
2'' we see that the derivative terms can be taken as in
(\ref{001}) with all generality and in particular, that the number
of internal components in $\au_n$ and $\au_{2N-n+1}$ (in the even
case) and in $\au_n$ and $\au_{2N-n}$ (in the odd one) are the
same.

%\bibliography{biblio}

\begin{thebibliography}{7}
\expandafter\ifx\csname
natexlab\endcsname\relax\def\natexlab#1{#1}\fi
\expandafter\ifx\csname bibnamefont\endcsname\relax
  \def\bibnamefont#1{#1}\fi
\expandafter\ifx\csname bibfnamefont\endcsname\relax
  \def\bibfnamefont#1{#1}\fi
\expandafter\ifx\csname citenamefont\endcsname\relax
  \def\citenamefont#1{#1}\fi
\expandafter\ifx\csname url\endcsname\relax
  \def\url#1{\texttt{#1}}\fi
\expandafter\ifx\csname
urlprefix\endcsname\relax\def\urlprefix{URL }\fi
\providecommand{\bibinfo}[2]{#2}
\providecommand{\eprint}[2][]{\url{#2}}

\bibitem[{\citenamefont{Barnich et~al.}(1994)\citenamefont{Barnich, Henneaux,
  and Tatar}}]{Barnich:1994pa}
\bibinfo{author}{\bibfnamefont{G.}~\bibnamefont{Barnich}},
  \bibinfo{author}{\bibfnamefont{M.}~\bibnamefont{Henneaux}}, \bibnamefont{and}
  \bibinfo{author}{\bibfnamefont{R.}~\bibnamefont{Tatar}},
  \bibinfo{journal}{Int. J. Mod. Phys.} \textbf{\bibinfo{volume}{D3}},
  \bibinfo{pages}{139} (\bibinfo{year}{1994}).

\bibitem[{\citenamefont{Henneaux}(1997)}]{Henneaux:1997bm}
\bibinfo{author}{\bibfnamefont{M.}~\bibnamefont{Henneaux}}
  (\bibinfo{year}{1997}), \eprint[http://arXiv.org/abs]{hep-th/9712226}.

\bibitem[{\citenamefont{Fernando Barbero~G and
  Villase\~nor}(2001)}]{Barbero:2000sb}
\bibinfo{author}{\bibfnamefont{J.}~\bibnamefont{Fernando Barbero~G}}
  \bibnamefont{and} \bibinfo{author}{\bibfnamefont{E.~J.~S.}
  \bibnamefont{Villase\~nor}}, \bibinfo{journal}{Nucl. Phys.}
  \textbf{\bibinfo{volume}{B600}}, \bibinfo{pages}{423} (\bibinfo{year}{2001}).

\bibitem[{\citenamefont{Barbero and Villase\~nor}(2000)}]{Barbero:1999ts}
\bibinfo{author}{\bibfnamefont{J.~F.} \bibnamefont{Barbero}} \bibnamefont{and}
  \bibinfo{author}{\bibfnamefont{E.~J.~S.} \bibnamefont{Villase\~nor}},
  \bibinfo{journal}{Phys. Rev.} \textbf{\bibinfo{volume}{D61}},
  \bibinfo{pages}{104014} (\bibinfo{year}{2000}).

\bibitem[{\citenamefont{Schwarz}(2000)}]{Schwarz:2000cb}
\bibinfo{author}{\bibfnamefont{A.}~\bibnamefont{Schwarz}}
  (\bibinfo{year}{2000}), \eprint[http://arXiv.org/abs]{hep-th/0011260}.

\bibitem[{\citenamefont{Barnich and Henneaux}(1993)}]{Barnich:1993vg}
\bibinfo{author}{\bibfnamefont{G.}~\bibnamefont{Barnich}} \bibnamefont{and}
  \bibinfo{author}{\bibfnamefont{M.}~\bibnamefont{Henneaux}},
  \bibinfo{journal}{Phys. Lett.} \textbf{\bibinfo{volume}{B311}},
  \bibinfo{pages}{123} (\bibinfo{year}{1993}).

\bibitem[{\citenamefont{Cattaneo et~al.}(1998)\citenamefont{Cattaneo, P.,
  Fucito, Martellini, M., Tanzini, and Zeni}}]{Cattaneo:1997fg}
\bibinfo{author}{\bibfnamefont{A.~S.} \bibnamefont{Cattaneo}},
  \bibinfo{author}{\bibfnamefont{C.-R.} \bibnamefont{P.}},
  \bibinfo{author}{\bibfnamefont{F.}~\bibnamefont{Fucito}},
  \bibinfo{author}{\bibfnamefont{M.}~\bibnamefont{Martellini}},
  \bibinfo{author}{\bibfnamefont{R.}~\bibnamefont{M.}},
  \bibinfo{author}{\bibfnamefont{A.}~\bibnamefont{Tanzini}}, \bibnamefont{and}
  \bibinfo{author}{\bibfnamefont{M.}~\bibnamefont{Zeni}},
  \bibinfo{journal}{Commun. Math. Phys.} \textbf{\bibinfo{volume}{197}},
  \bibinfo{pages}{571} (\bibinfo{year}{1998}).

\end{thebibliography}

\end{document}